\documentstyle [preprint,prl,aps] {revtex}
\begin{document}
\draft
\parskip = 3mm
\begin{titlepage}
\preprint{\vbox{\hbox{IASSNS-AST 94/57}}}
\title{ \large \bf 
Limits on Electron-Neutrino Oscillations from \\
the GALLEX $^{51}$Cr
Source Experiment}
\author{J.N. Bahcall, P.I. Krastev\footnote{Permanent address: 
Institute for Nuclear Research and Nuclear Energy, Bulgarian Academy
of Sciences, Sofia, Bulgaria} and E. Lisi\footnote{Dipartimento di
Fisica and Sezione INFN di Bari, Bari, Italy}}
\address{Institute for Advanced Study, Princeton, NJ 08540}

\maketitle 
\begin{abstract}
The recent result from the chromium source experiment carried out by
the GALLEX collaboration implies interesting limits on the parameters
$\Delta m^2$ and $\sin^22\theta$ describing neutrino oscillations.
Values of $\Delta m^2~>~0.17$ eV$^2$ for maximal mixing and of
$\sin^22\theta > 0.38$ for $\Delta m^2 > 1$~eV$^2$ are ruled out at
90\% C.L. This result improves by more than an order of magnitude
previous limits on $\Delta m^2$ derived from electron-neutrino
oscillation experiments at accelerators.
\end{abstract} 

\end{titlepage} 
\newpage 

The GALLEX collaboration has announced recently the first result of
their experiment with an artificial neutrino source \cite{calib}. This
is the first successful test of a solar neutrino experiment with an
artificial radioactive source of electron neutrinos. In this note we
use the fact that the $^{71}$Ge production rate reported by GALLEX is
close to the expected rate to derive limits on the electron-neutrino
oscillation parameters.

The GALLEX experiment is described in detail in \cite{calib,nim}. The
neutrinos emitted from the radioactive source, $^{51}$Cr, are
monoenergetic with energies 0.746 (81\%), 0.751 (9\%), 0.426 (9\%) and
0.431 (1\%) MeV. The target solution, GaCl$_3$, fills a cylindrical
container with radius 1.9 m to a height of 5 m. The source is located
in a cavity, approximately 2.3 m below the surface of the GaCl$_3$.

The production rate, $Q_{\rm Ge}$, of $^{71}$Ge atoms in the absence of
oscillations is given by the following integral over the volume, $V$,
of the detector:
\begin{equation}
Q_{\rm Ge} = \int\limits_V \sum_{i=1,4}
\frac{\Phi_i(E_{\nu}^i)}{4\pi {\rm r}^2} N_{\rm Ga}
\sigma(E_{\nu}^i)dV\quad,
\end{equation}
where $\Phi_i$ is the rate of neutrino emission in each $^{51}$Cr
decay mode, $r$ is the distance the neutrinos travel between
production and capture, $N_{\rm Ga}$ is the number density of
$^{71}$Ga atoms, $\sigma$ is the neutrino capture cross-section, and
$E_{\nu}^i$ is the neutrino energy. We have computed $Q_{\rm Ge}$
taking into account the detailed geometry of the detector. Our result
agrees with the estimate given in \cite{calib} with an accuracy of
better than 1\%.

The GALLEX collaboration reports a result of 1.04 $\pm$ 0.12 for the
ratio, $R^{exp}$ = $Q_{\rm Ge}^{exp}$/$Q^{\phantom{exp}}_{\rm Ge}$, of
the measured production rate of $^{71}$Ge atoms to the rate predicted
using the cross-section calculated in \cite{gacr}. Including the
estimated theoretical error \cite{gacr} of $\pm 3$\% gives a total
error of $\pm 0.13$.

If the electron neutrino produced in a $^{51}$Cr decay oscillates
\cite{osc} into another neutrino type, $\nu_{\mu}, \nu_{\tau}$ or a
sterile neutrino, $\nu_s$, the $^{71}$Ge production rate,
$Q^{osc}_{\rm Ge}$, in the target solution will be reduced.  To
calculate $Q_{\rm Ge}^{osc}$ one must convolve the integrand in
Eq.~(1) with the well known neutrino survival probability in vacuum.
The estimated ratio $R^{osc}$ = $Q_{\rm
Ge}^{osc}/Q^{\phantom{osc}}_{\rm Ge}$ depends on the parameters
$\Delta m^2$ and $\sin^22\theta$ which determine the neutrino survival
probability. The requirement that $R^{osc}$ does not differ
significantly from $R^{exp}$ constrains the allowed region for these
parameters. Since $R^{exp}$ lies slightly above the physical region
for $R^{osc}$ ($0 \leq R^{osc} \leq 1$), we renormalize the
distribution of the total error using a Bayesian approach with a flat
prior distribution, as described in \cite{pdb}.

The results of our analysis are shown in Fig.~1. Values above and to
the right of the full curve are ruled out at 90\% C.L. In particular,
values of $\Delta m^2 > 0.17$ eV$^2$ for $\sin^22\theta = 1$ and of
$\sin^22\theta > 0.38$ for $\Delta m^2 >$ 1 eV$^2$ are excluded by
this analysis. For a C.L. of 95\% the unacceptable range is only
slightly contracted to $\Delta m^2 > 0.19$ eV$^2$ for $\sin^22\theta =
1$ and to $\sin^22\theta > 0.45$ for $\Delta m^2 > 1$ eV$^2$.

Our result improves by more than an order of magnitude previous limits
on the maximum $\Delta m^2$ allowed by accelerator experiments. The
best upper limit from electron-neutrino disappearance experiments
\cite{old1,enriq} is $\Delta m^2 < 2.3$ eV$^2$ for $\sin^22\theta =
1$. In the $\nu_e\rightarrow\nu_{\tau}$ appearance experiment
\cite{ushida} a weaker limit, $\Delta m^2 < 9$ eV$^2$, has been
obtained.  The reason the GALLEX experiment implies a better limit on
$\Delta m^2$, despite the small distance between source and target, is
that the energy of the neutrinos from $^{51}$Cr decay ($<$ 1 MeV) is
much lower than the typical neutrino energy ($\sim 30$ MeV to $\sim
50$ GeV) in accelerator experiments.

Our upper limit on the mixing angle for large $\Delta m^2$ is better
than was obtained in \cite{old1} but does not improve the best
existing limits from Refs.~\cite{enriq,ushida}. The most stringent
limit from accelerator neutrino experiments is $\sin^22\theta <
7\times 10^{-2}$
\cite{enriq}, which however is reached only for values of $\Delta m^2
> 100$ eV$^2$.

We note in passing that the gallium solar neutrino experiments also
place a strong limit on possible electron charge non-conserving
interactions. Since $^{71}$Ga is heavier than $^{71}$Ge, the decay
$^{71}$Ga $\rightarrow$ $^{71}$Ge is only forbidden by charge
conservation. From the observed rate in the gallium solar neutrino
experiments \cite{sagal}, one can conclude that the ratio of charge
non-conserving coupling constant, $\epsilon\, G_F$, divided by the
usual weak interaction coupling constant, $G_F$, is very small
\begin{equation}
\epsilon < 7\times 10^{-14} \quad ,
\end{equation}
if the charge non-conserving interaction is also described by a
four-fermion interaction (see the second of Refs.~\cite{gacr},
p. 359).

In conclusion, the limits on neutrino oscillations implied by the
recently announced GALLEX radioactive source experiment improve
thirteen-year old limits obtained in electron-neutrino oscillation
experiments at accelerators.

\smallskip
{\bf Acknowledgments.} We are grateful to the GALLEX collaboration for
sending us a preprint of their important paper, and especially to
K. Rowley for a precise sketch of the experimental geometry. J.N.B.
acknowledges support from NSF grant \#PHY 92-45317.  The work of
P.I.K.  was partially supported by Dyson Visiting Professor Funds from
the Institute for Advanced Study and the work of E.L. was supported by
a post-doctoral INFN fellowship.

\centerline{\bf Figure Captions} 
\vspace*{1.0 cm} 

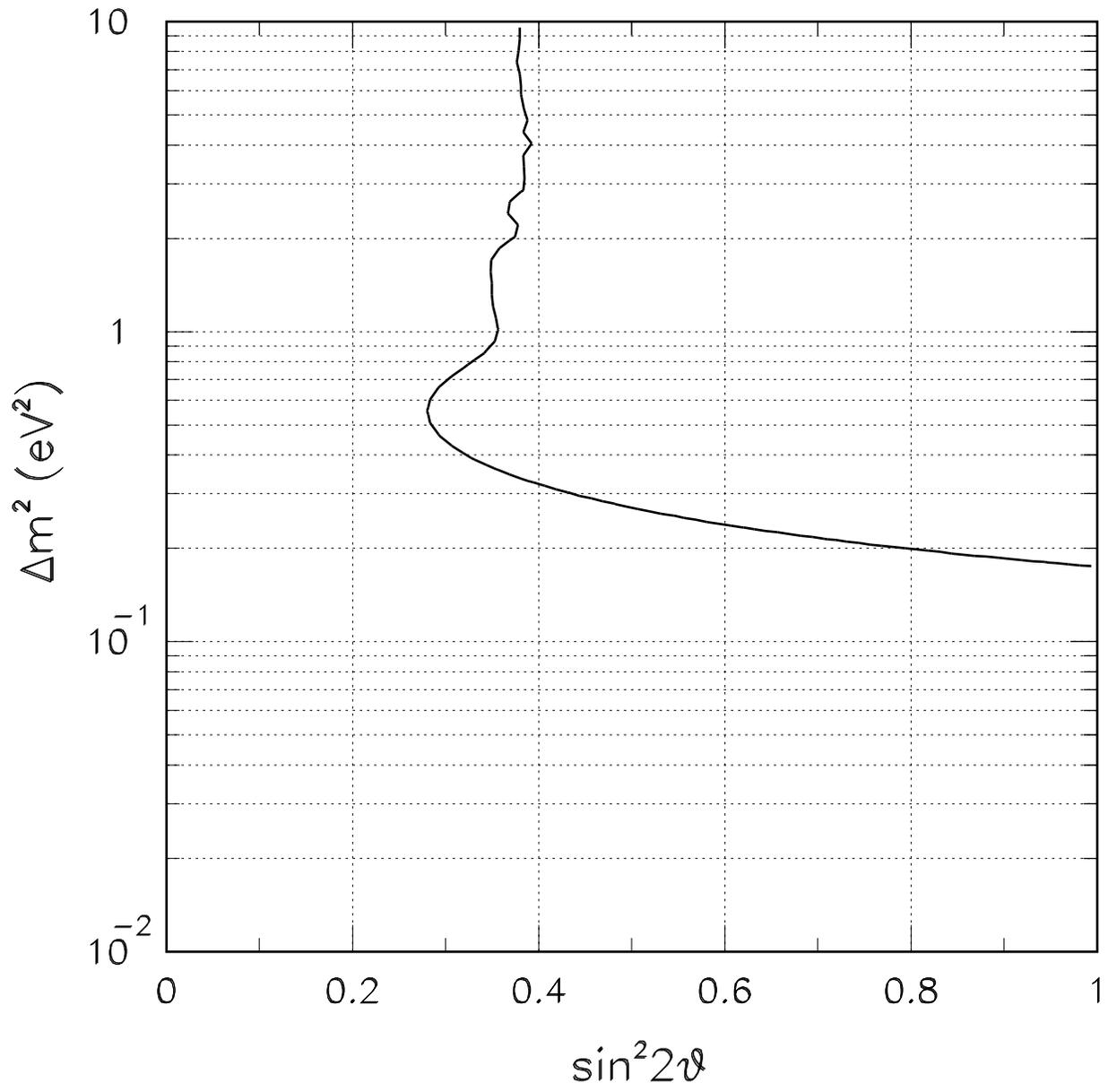
\noindent {\bf Fig.~1.} Region of electron neutrino 
oscillation parameters ruled out at 90\% C.L. by the GALLEX $^{51}$Cr
source experiment.

\end{document}